\pdfoutput=1

\documentclass[fleqn,10pt]{wlscirep}

\usepackage[utf8]{inputenc}
\usepackage[T1]{fontenc}
\usepackage{trace}
\usepackage{color}
\usepackage{amsmath}

\title{Deep Neural Network Inverse Design of Integrated Nanophotonic Devices}

\author[1]{Mohammad H. Tahersima}
\author[1,*]{Keisuke Kojima}
\author[1]{Toshiaki Koike-Akino}
\author[1]{Devesh Jha}
\author[1]{Bingnan Wang}
\author[1]{Chungwei Lin}
\author[1]{Kieran Parsons}

\affil[1]{Mitsubishi Electric Research Laboratories, 201 Broadway, Cambridge, MA 02139, USA}
\affil[*]{kojima@merl.com}


\begin{abstract}
Predicting physical response of an artificially structured material is of particular interest for scientific and engineering applications. 
Here we use deep learning to predict optical response of artificially engineered nanophotonic devices. 
In addition to predicting forward approximation of transmission response for any given topology, this approach allows us to inversely approximate designs for a targeted optical response. 
Our Deep Neural Network (DNN) could design compact ($2.6 \times 2.6~\mu \textrm{m}^2$) silicon-on-insulator (SOI)-based $1 \times 2$ power splitters with various target splitting ratios in a fraction of a second. 
This model is trained to minimize the reflection (smaller than $20$~dB) while achieving maximum transmission efficiency (above $90 \%$) and target splitting specifications. 
This approach paves the way for rapid design of integrated photonic components relying on complex nanostructures.
\end{abstract}

\begin{document}
\flushbottom
\maketitle
\thispagestyle{empty}

\section*{Introduction}
Artificially engineered subwavelength nanostructured materials can be used to control incident electromagnetic fields into specific transmitted and reflected wavefronts. 
Recent nanophotonic devices have used such complex structures to enable novel applications in optics, integrated photonics, sensing, and computational metamaterials in a compact and energy-efficient form \cite{ni2015ultrathin, alu2005achieving, monticone2013full, arbabi2016multiwavelength, khorasaninejad2016metalenses, krasnok2017nonlinear, azad2016metasurface, lalau2013adjoint, motayed2018highly, silva2014performing}. 
Nevertheless, optimization of nanostructures, with enormous number of possible combination of features, using numerical simulation is computationally costly.
For example, computing electromagnetic field profile via finite-difference time-domain (FDTD) methods may require long simulation time, several minutes to hours depending on the area of photonic device, for analyzing the optical transmission response.
In order to design nanostructures achieving target transmission profile, we need to perform a large number of FDTD simulations in most meta-heuristic approaches.
To resolve the issue, we previously developed an artificial intelligence integrated optimization process using neural networks (NN) that can accelerate optimization by reducing required number of numerical simulations \cite{kojima2017acceleration, teng2018broadband} to demonstrate how NNs can help to streamline the design process.

\begin{figure}[ht]
\centering
\includegraphics[width=14cm]{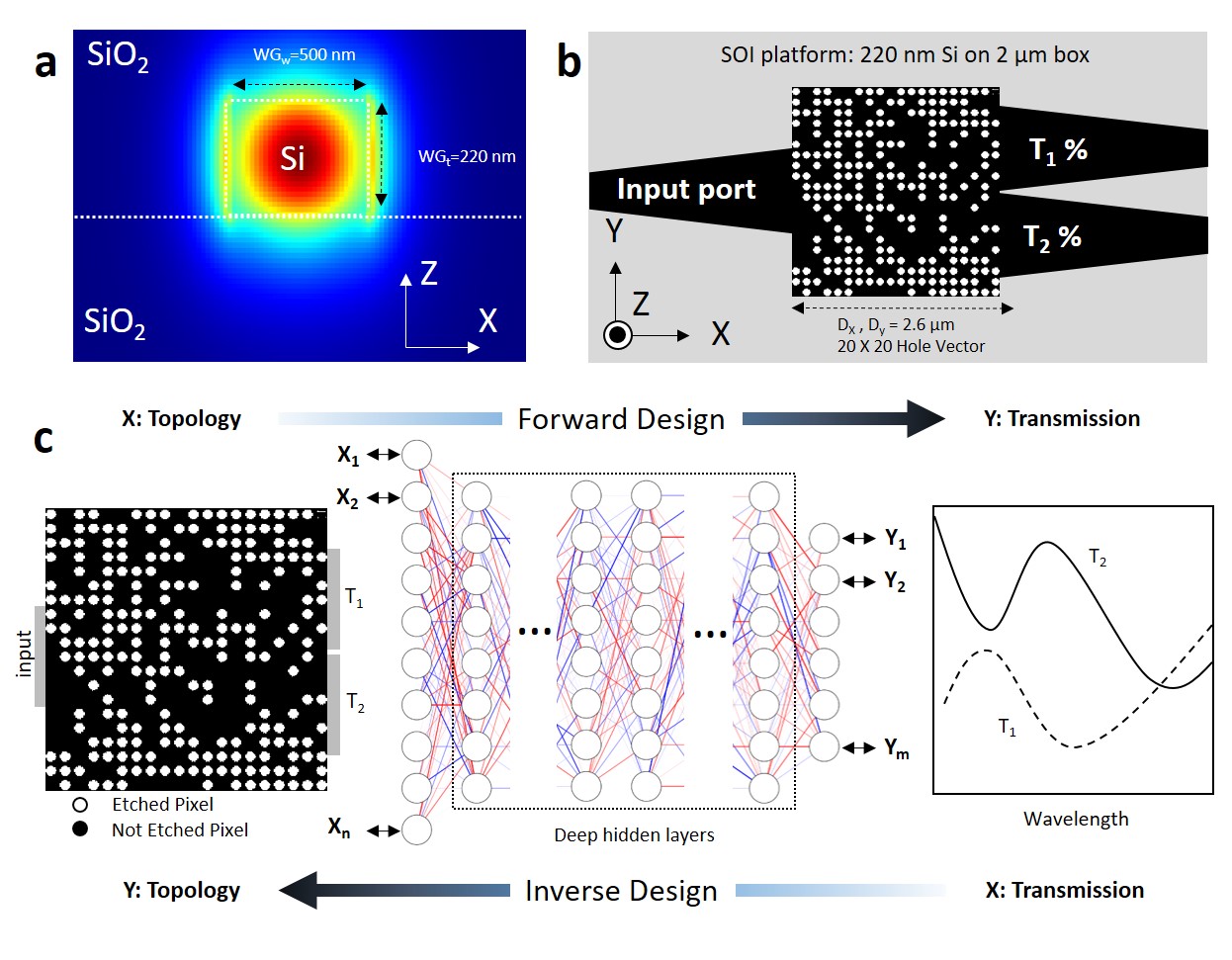}
\caption{ a) TE mode is launched into the standard SOI waveguide at input port of power splitter.
 b) Schematic of standard SOI-based MMI power splitter, whose size is chosen to be $2.6 \times 2.6 \mu \textrm{m}^2$.
 Circles indicates etch holes to be optimized.
 By optimizing binary sequence of position of etch holes it is possible to adjust light propagation into either of the ports.
 c) We use DNN for forward and inverse modeling of nanophotonic devices.
 The DNN can take device topology design as input and spectral response of the metadevice as label or vice versa.}
\label{fig:Figure1}
\end{figure}

Deep learning methods are representation-learning techniques obtained by composition of non-linear models that transform the representation at the previous level into a higher and slighly more abstract level in a hierarchical manner~\cite{lecun2015deep}. The main idea is that by cascading a large number of such transformations, very complex functions can be learnt in a data-driven fashion using deep neural networks~\cite{krizhevsky2012imagenet}. The huge sucess of deep learning in modeling complex input-ouput relationship has attracted attention from several scientific communities such as material discovery\cite{ghaboussi1991knowledge}, high energy physics \cite{baldi2014searching}, single molecule imaging {yasui2018automated}, medical diagnosis \cite{jun2018deep}, and particle physics \cite{radovic2018machine}. It has received some attention in optical community and there has been some recent work on reverse modeling for design of nano-structured optical components using DNN \cite{liu2018training, ma2018deep, malkiel2018deep, peurifoy2018nanophotonic,sun2018efficient, asano2018optimization}, as well as optical implementation of an artificial neural network \cite{tait2017neuromorphic, mehrabian2018pcnna, lin2018all, chiles2018design}.
NNs can be used to predict a desired optical response of a topology (Forward Design) as well as design a topology given a desired optical response (Inverse Design). 

Inverse design of photonic structures were conventionally demonstrated using adjoint sensitivity analysis \cite{piggott2015inverse, piggott2017fabrication, frandsen2016inverse, cao2003adjoint}. 
More recently, D. Liu used a tandem NN architecture to learn non-unique electromagnetic scattering of alternating dielectric thin films with varying thickness \cite{liu2018training}. 
J. Peurifoy demonstrated NNs to approximate light scattering of multilayer shell nanoparticles of SiO\textsubscript{2} and TiO\textsubscript{2} using a fully connected NNs with a depth of $4$ layers having $100$ neurons \cite{peurifoy2018nanophotonic}.
During preparation of this paper, T. Asano provided a neural network for prediction of the quality factor in two dimensional photonic crystals \cite{asano2018optimization}.
Inspired by this progress, we aim to train a NN that can instantaneously design an integrated photonic power divider with a ratio specified by the user. 
The design space for integrated photonic device is considerably larger than previously demonstrated optical scattering applications, that call for robust deeper networks such as Deep Residual Networks (ResNet)\cite{he2016deep}.

Integrated photonic beam splitters based on a multimode interferance (MMI) have been widely used to equally divide the power into the output ports. 
Although an arbitrary split ratio can be applied in various applications such as signal monitoring, feedback circuits, or optical quantization \cite{kang2014resolution}, the design space is hardly explored due to design complexity. 
Tian et al.\cite{tian2018broadband} demonstrated SOI-based $1 \times 3$ coupler with variable splitting ratio in a $15 \times 15~\mu \textrm{m}^2$ device footprint with $60$ nm bandwidth and $80 \%$ transmission efficiency.
Xu et al.\cite{xu2017integrated} optimized positioning of squared etched pixels to achieve $80 \%$ efficiency for arbitrary ratio power dividers in $3.6 \times 3.6~\mu \textrm{m}^2$ device foot print.

To design photonic power divider with arbitrary splitting ratio, the designer often begins with an overall structure based on analytical models and fine tune the structure using parameter sweep in numerical simulations. 
Here, we demonstrate that using deep learning methods we could efficiently learn the design space of a broadband integrated photonic power divider in a compact residual neural network model. 
This method allows design by specifications where user simply asks for a specific power splitting performance and can see the near ideal solution almost instantaneously without depending on time-consuming FDTD simulations. 
Our device has above 90$\%$ transmission efficiency in a footprint of $2.6 \times 2.6~\mu \textrm{m}^2$, which to the best of our knowledge, is the smallest arbitrary ratio beam splitter to date.
Moreover, our design does not rely on arbitrary device morphologies and is constrained to a $20 \times 20$ vector of etched holes with a radius of $45$~nm, conveniently fabricatable by the current semiconductor technology.

\section*{Deep Learning for Forward Modeling to Predict Optical Response}
\subsection*{Simulation Setting and Dataset}
When a broadband light encounters an obstacle, with a different refractive index, along its path, it undergoes reflection, refraction, and scattering. 
The goal of nanostructured integrated photonics power splitter is to organize the optical interaction events, such that the overall effect of the ensemble of scattering evets guides the beam to a target port and power intensity. 
To design the power ratio splitter using DNN we chose a simple three port structure on a standard fully etched SOI platform. 
One input and two outputs $0.5 \mu \textrm{m}$ wide port are connected using an adiabatic taper to the $2.6 \mu \textrm{m}$ wide square power splitter design region with a connection width of $1.3 \mu \textrm{m}$ (Figure~\ref{fig:Figure1}).  
We use numerical simulation (Methods section) to generate labeled data for training the network. 
We then feed the DNN with numerical optical experiments and train a neural network able to represent the relationship between hole vectors and spectral response at each port.   
Initially our input data are several $20 \times 20$ hole vectors (HV), each labeled by its spectral transmission response (SPEC) at port 1 and 2 and reflection from the input port.
Each pixel is a circle with a radius of $45$~nm that is easily fabricable using conventional lithography methods \cite{lu2016ultra, piggott2017fabrication}.
Each pixel can have a binary state of 1 for etched ($n = n_\mathrm{Silicon}$) and 0 for not etched ($n = n_\mathrm{Silica}$) (See Methods).
Changing the refractive index at a hole position modifies the local effective index inside of the power divider to determine the propagation path for the travelling wave in the device. 
We use random and carefully chosen patterned initial HVs and optimize them using heuristic optimization approaches for various optimization metrics to collect a diverse set of labeled training data for supervised learning.  

\begin{figure}[ht]
\centering
\includegraphics[width=12cm]{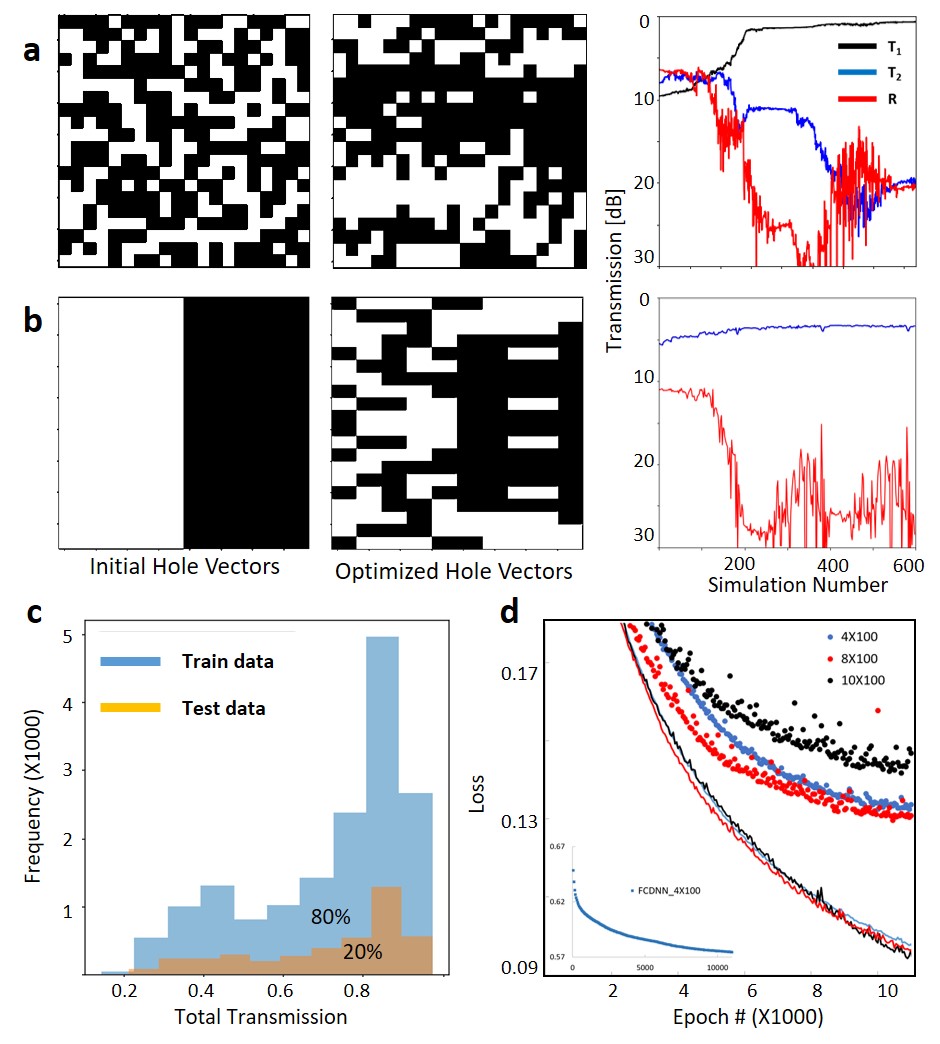}
\caption{We train the network with a diverse set of data, each starting with an initial condition, etched hole density, and various metrics to optimize a spectral response. 
We generate approximately $20{,}000$ etched hole vector data with transmission labels. 
Here we show two set of these data: a) asymmetric search optimization to maximize $T_1$ with an random initial vector, b) symmetric search to maximize $\min(T_1) + \min(T_2)- \alpha \times \max(|R|)$ ($\alpha = 4$) with a patterned initial vector. 
c) Histogram of all transmission train and test data labels collected by numerical methods for $20{,}000$ power splitter topologies at $1550$ nm.
 d) Learning curve for $\sim 10{,}000$ epochs of training. Learning curve for deep residual neural network shows the network loss reduces by increasing depth of network up to $8$ layers. The inset shows the best case for a FCDNN that has significantly larger loss value ($\sim 0.58$). Here all cost functions are based on negative log likelihood}
\label{fig:Figure2}
\end{figure}

For the forward problem, the input is the two-dimensional array ($20 \times 20$) which corresponds to the binary image for the hole locations. We train a DNN to predict the transmission and reflection spectra vector which is $63$-dimensional. The $63$-dimensional vector includes spectral data ($1450$ to $1650$~nm) for transmission at both port 1, port 2, and reflection at the input port. For the inverse design,the $63$-dimensional vector is used as an input and the hole vectors are considered as the labels. Thus, the forward problem is solved as a regression problem while the inverse problem is solved as a classification problem, where we predict a binary vector representing the hole locations. For both the problems, we first used a fully-connected DNN (FCDNN) with mulitple layers where each layer has $100$ neurons. The number of layers was considered as a hyperparameter which was optimized during the numerical experiments. However, we found that increasing the depth of the fully-connected DNNs didn't improve the performance of the network. Conseuently, we used a residual deep neural network (ResNet) to train both the forward and inverse problem (see Figure~\ref{fig:Figure2}d for a qualitative comparison between regular DNNs and ResNets). ResNets have emperically been proven to allow more flexibility in training deep architectures than the conventional neural networks~\cite{he2016deep}. The main idea is that the ResNet uses an additional identity function to allow smooth forward and backward propagation of gradients~\cite{he2016deep}. A Gaussian log-likelihood function is used to train the DNN modeling the forward design problem. A Bernoulli log-likelihood classifier is used as the loss function for training the inverse problem. The Gaussian log-likelihood loss function is represented by the following equation. 
\begin{align}
 -\log P(Y|X,\mathbf{W}) &=
 \frac{1}{K} \sum_{n}^K
 \left(
 \frac{1}{2}\log\left( 2\pi \sigma^2 \right)
 + \frac{1}{2 \sigma^2} (y_n - \mathbf{W}^Tx_n)^2
 \right)
 ,
\end{align}
where $P(Y|X,\mathbf{W})$ denotes the probabilistic model, $\mathbf{W}$ denotes the model parameters $K$ is the number of training data. The loss function is optimized using the Adam optimization algorithm~\cite{kingma2014adam}. Training is terminated after a fixed number of iterations to ensure convergence. The training and validation results were very similar for our trained networks and thus we didn't use any regularization for over-fitting.

\begin{figure}[ht]
\centering
\includegraphics[width=16cm]{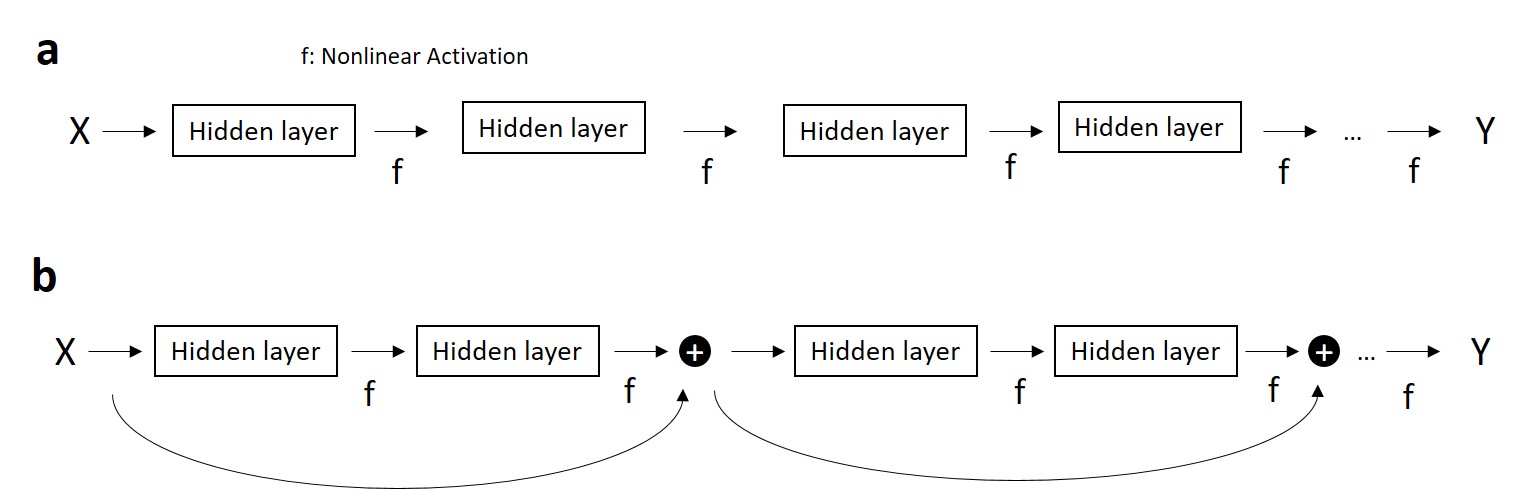}
\caption{Network architecture for plain DNN (a) and ResNet (b) used for inverse design of integrated nanophotonics. 
We use Sigmoid activation function in both network architectures. 
ResNet architecture can achieve accuracy from increase network depth}
\label{fig:Figure3}
\end{figure}

\subsection*{Results}
To test the nanostructured power divider network, first we use a randomly selected, unseen $20 \%$ data from the same set of simulation data used to train the network. 
The test data set helps to prevent overfitting the model to the training data (Figure~\ref{fig:Figure2}d).

\begin{figure}[ht]
\centering
\includegraphics[width=16cm]{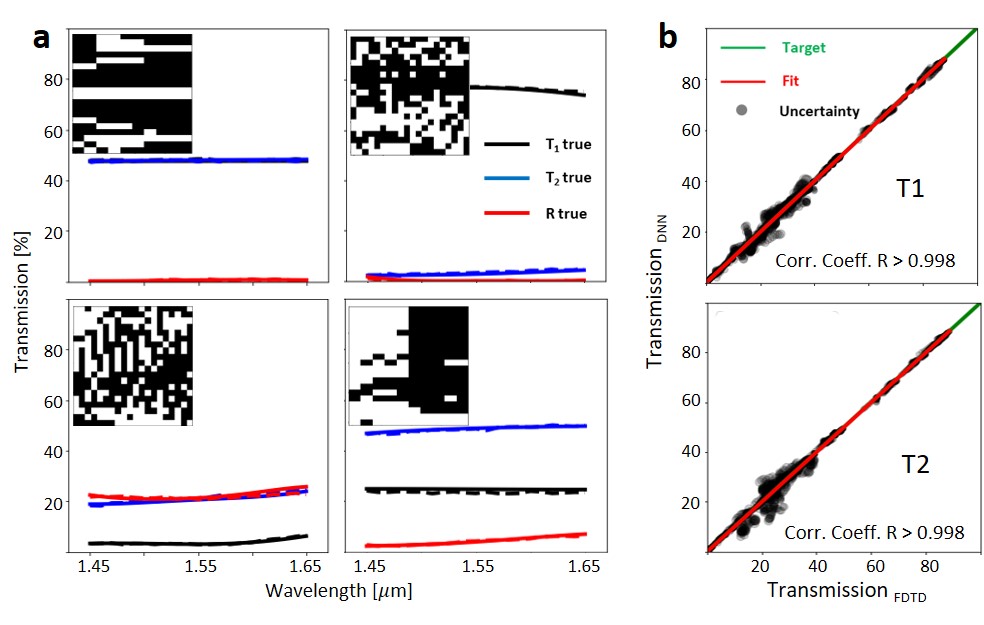}
\caption{Spectrum approximation using deep ResNet. 
$8{,}000$ input data were used for training and $2{,}000$ data were used for testing. 
a) Comparison of predicted spectral response of power splitter using ResNet to the true value of the spectra shown for $6$ cases. 
Black, blue, and red colors stand for transmission at port 1, transmission at port 2, and reflection at the input port, respectively. 
Solid lines are true values for a given hole vector and dashed lines are predicted spectral response using ResNet. 
b) Fitting ResNet predicted transmission values versus true transmission values at each output port. 
The correlation coefficient $R$ is above $0.998$ across the full range of transmission ratios (0 to 1) and approaches unity as transmission increases. 
Gray circle symbol size is proportional to gradient uncertainty.}
\label{fig:Figure4}
\end{figure}

\begin{figure}[ht]
\centering
\includegraphics[width=\linewidth]{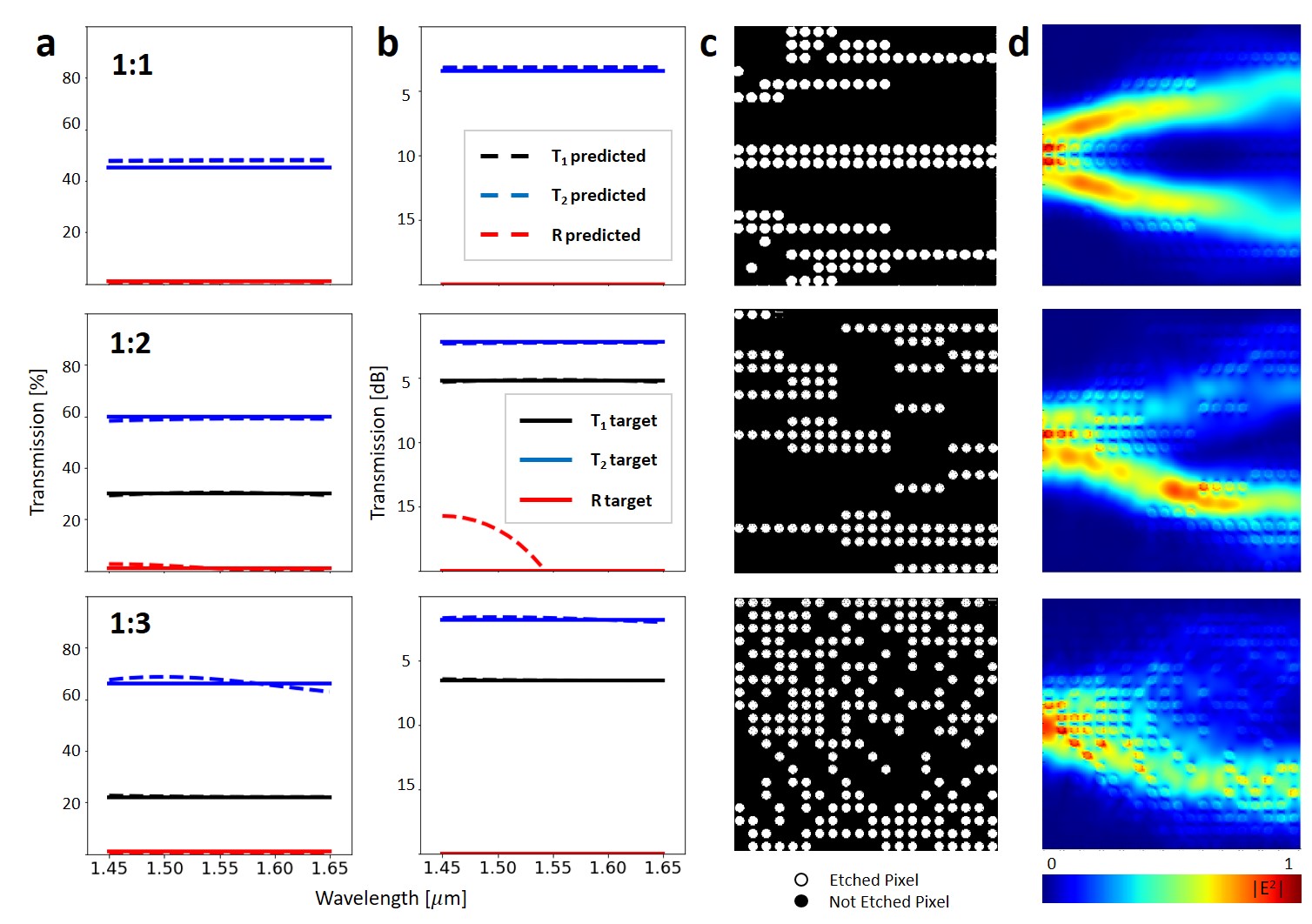}
\caption{Demonstration of inverse design using ResNet for $90 \%$ efficient power splitter with splitting ratios of 1:1, 1:2, 1:3. 
a,b) spectral response for transmission at port 1, 2, and reflection to the input port. 
Solid lines are target broadband response from the power splitter and match well with dashed lines which are simulated values for predicted hole vector of the target response. 
c) Predicted hole vector topology for target spectral response (solid line in a, b). 
d) Simulated optical power propagation through DNN predicted power splitter}
\label{fig:Figure5}
\end{figure}

In the following we present the outcome of the network for forward prediction of spectra from HV (Figure~\ref{fig:Figure4}), and inversely designing the HV from a given physical feasible SPEC specifications (Figure~\ref{fig:Figure5}). 
First we test the forward computation of the network to see prediction of spectral response of a topology that the network is not trained on. Interestingly, the network could predict transmission and reflection spectra quite accurately (\ref{fig:Figure4}a).
Our design goal is to achieve compact naostructured power splitters with high transmission efficiency and minimized back reflection. Low back reflection is of great importance since in active integrated photonic circuits, it is important to minimize the back reflection. That is why we use reflection factors larger than $2$, in our optimization metrics, to emphasize minimizing back reflection in these power splitters.  

To quantify the prediction accuracy we use a correlation plot that compares true numerically verified optical response with DNN predictions. The correlation coefficient of the DNN prediction was above $99\%$ (\ref{fig:Figure4}b). 
We use the variance of the negative log likelihood cost function as a mean to determine the confidence of the neural network and show it as the area of the prediction uncertainty in the correlation plot. 
We observe that confidence of the prediction is lower in lower transmission regime and improves at higher transmission regimes. 
This is expected since the training data is mainly contain high transmission devices (\ref{fig:Figure2}d).      

We test the inverse modeling on the same data as above by using SPECs as data and HVs as label and reversing and optimizing the inverse network. 
To test the generalization capabilities of the network, we investigate the network’s inverse design performance on arbitrary and unfamiliar cases. 
To do this, we generate a reference table containing broadband constant transmission values for each port and use them as the input data batch for the Inverse Design DNN model. 
The predicted HVs can take any value from 0 to 1 from a Bernoulli distribution classifier. The classification converges to 0 or 1 as the loss reduces by increasing the number of training epochs. The quantized binary sequence is then fed back into the numerical solver. 
In a next step, we run independent FDTD simulation to check validity of the response (Figure~\ref{fig:Figure5}). 
Numerically simulated electric field propagation at center wavelength of 1550 nm for three splitting ratios of 1:1, 1:2, and 1:3 show various power splitting mechanisms from classical MMI based beam splitters. 
A symmetric electric field distribution intensity can be observed in the case of 1:1 splitting. On the other hand the electric field intensity is asymmetric for asymmetric splitting ratio devices (as expected) and beam path is broadened for high output side $T_2$. 

In all these three cases the transmission efficiency exceeds $90 \%$ which, to the best of our knowledge, is the highest transmission efficiency demonstrated in integrated power splitters. And this is also the first time minimizing reflection is taken into consideration.
Although we did not aim to maximize the operation bandwidth as an objective, our power splitters show broadband transmission between $1450$~nm to $1650$~nm.
Additionally, we set the reflection target at -20 dB. We achieved the reflection less than -20 dB, except for the 1:2 spliter case.
 
\subsection*{Discussion}

NNs can be used to take device structure data (shape, depth, and permittivity) to predict the optical response of the nanostructure (forward network). 
In this case NN can be used as method for fast approximation of the optical response instead use of computationally heavy numerical methods. 
Another way to use NNs, which is not available in numerical method, is taking an optical response and providing the user with an approximate solution of nanostructure (inverse design). 
Although DNN initially need a large amount of data set for the training purpose, it is possible to process several heuristic optimization metrics in parallel on a computing cluster to speed up generating the training data. 
Once the network is trained to represent the topology as optical response and vice versa, it can design the nanostructured geometry in a fraction of second.

 In conclusion, we demonstrated application of DNNs in design of nanostuctured integrated photonic components.
Although the design space for this problem is very large ($2^{400}$) possible combinations), by training DNN with nearly $20{,}000$ simulation data we trained a network that can approximate the spectral response of the an arbitrary hole vector within this design space. 
In addition, we could use the inverse network to design a nearly optimized power splitter topology for any user specific spectral responses.   
The capability of DNN in predicting optical response of a topology and the inverse design holds promise in wide spread use of these networks in design of nanostructured photonic systems.

\section*{Methods}
\subsection*{Numerical Simulations}
We use Lumerical’s FDTD simulation package for generating the training data. 
The data contains more than $20{,}000$ numerical simulations, where each experiment is a 3D FDTD simulation composed of passive SOI waveguides and the beam splitter device. 
Initial random hole position matrix of the beam splitter was generated, exported, and manipulated using the MATLAB automation. 
Hole position generating script uses different algorithms (such as Dirct Binary Search), initial conditions, and optimization metrics to sequentially create a robust neural network representation of the device structure. 
Dispersive refractive indices of silicon and silica from literature\cite{palik1997hand} were used for all simulations for a broadband simulation in the range of $1.45 \mu \textrm{m} -- 1.65 \mu \textrm{m}$.
The fundamental TE mode at $1550$~nm was used at input source and TE mode output power was recorded for transmission and reflection. 
We note that TM mode output is below $10^{-5}$.

\subsection*{Deep Neural Network (DNN)}
We used the open source machine learning framework of Tensorflow in python language to build and test our deep neural networks.

\bibliography{main}

\begin{thebibliography}{10}
\urlstyle{rm}
\expandafter\ifx\csname url\endcsname\relax
  \def\url#1{\texttt{#1}}\fi
\expandafter\ifx\csname urlprefix\endcsname\relax\def\urlprefix{URL }\fi
\expandafter\ifx\csname doiprefix\endcsname\relax\def\doiprefix{DOI: }\fi
\providecommand{\bibinfo}[2]{#2}
\providecommand{\eprint}[2][]{\url{#2}}

\bibitem{ni2015ultrathin}
\bibinfo{author}{Ni, X.}, \bibinfo{author}{Wong, Z.~J.},
  \bibinfo{author}{Mrejen, M.}, \bibinfo{author}{Wang, Y.} \&
  \bibinfo{author}{Zhang, X.}
\newblock \bibinfo{journal}{\bibinfo{title}{An ultrathin invisibility skin
  cloak for visible light}}.
\newblock {\emph{\JournalTitle{Science}}} \textbf{\bibinfo{volume}{349}},
  \bibinfo{pages}{1310--1314} (\bibinfo{year}{2015}).

\bibitem{alu2005achieving}
\bibinfo{author}{Al{\`u}, A.} \& \bibinfo{author}{Engheta, N.}
\newblock \bibinfo{journal}{\bibinfo{title}{Achieving transparency with
  plasmonic and metamaterial coatings}}.
\newblock {\emph{\JournalTitle{Physical Review E}}}
  \textbf{\bibinfo{volume}{72}}, \bibinfo{pages}{016623}
  (\bibinfo{year}{2005}).

\bibitem{monticone2013full}
\bibinfo{author}{Monticone, F.}, \bibinfo{author}{Estakhri, N.~M.} \&
  \bibinfo{author}{Al\`u, A.}
\newblock \bibinfo{journal}{\bibinfo{title}{Full control of nanoscale optical
  transmission with a composite metascreen}}.
\newblock {\emph{\JournalTitle{Phys. Rev. Lett.}}}
  \textbf{\bibinfo{volume}{110}}, \bibinfo{pages}{203903},
  \doiprefix\url{10.1103/PhysRevLett.110.203903} (\bibinfo{year}{2013}).

\bibitem{arbabi2016multiwavelength}
\bibinfo{author}{Arbabi, E.}, \bibinfo{author}{Arbabi, A.},
  \bibinfo{author}{Kamali, S.~M.}, \bibinfo{author}{Horie, Y.} \&
  \bibinfo{author}{Faraon, A.}
\newblock \bibinfo{journal}{\bibinfo{title}{Multiwavelength
  polarization-insensitive lenses based on dielectric metasurfaces with
  meta-molecules}}.
\newblock {\emph{\JournalTitle{Optica}}} \textbf{\bibinfo{volume}{3}},
  \bibinfo{pages}{628--633} (\bibinfo{year}{2016}).

\bibitem{khorasaninejad2016metalenses}
\bibinfo{author}{Khorasaninejad, M.} \emph{et~al.}
\newblock \bibinfo{journal}{\bibinfo{title}{Metalenses at visible wavelengths:
  Diffraction-limited focusing and subwavelength resolution imaging}}.
\newblock {\emph{\JournalTitle{Science}}} \textbf{\bibinfo{volume}{352}},
  \bibinfo{pages}{1190--1194} (\bibinfo{year}{2016}).

\bibitem{krasnok2017nonlinear}
\bibinfo{author}{Krasnok, A.}, \bibinfo{author}{Tymchenko, M.} \&
  \bibinfo{author}{Alù, A.}
\newblock \bibinfo{journal}{\bibinfo{title}{Nonlinear metasurfaces: a paradigm
  shift in nonlinear optics}}.
\newblock {\emph{\JournalTitle{Materials Today}}}
  \textbf{\bibinfo{volume}{21}}, \bibinfo{pages}{8 -- 21},
  \doiprefix\url{https://doi.org/10.1016/j.mattod.2017.06.007}
  (\bibinfo{year}{2018}).

\bibitem{azad2016metasurface}
\bibinfo{author}{Azad, A.~K.} \emph{et~al.}
\newblock \bibinfo{journal}{\bibinfo{title}{Metasurface broadband solar
  absorber}}.
\newblock {\emph{\JournalTitle{Scientific Reports}}}
  \textbf{\bibinfo{volume}{6}} (\bibinfo{year}{2016}).

\bibitem{lalau2013adjoint}
\bibinfo{author}{Lalau-Keraly, C.~M.}, \bibinfo{author}{Bhargava, S.},
  \bibinfo{author}{Miller, O.~D.} \& \bibinfo{author}{Yablonovitch, E.}
\newblock \bibinfo{journal}{\bibinfo{title}{Adjoint shape optimization applied
  to electromagnetic design}}.
\newblock {\emph{\JournalTitle{Optics express}}} \textbf{\bibinfo{volume}{21}},
  \bibinfo{pages}{21693--21701} (\bibinfo{year}{2013}).

\bibitem{motayed2018highly}
\bibinfo{author}{Motayed, A.} \emph{et~al.}
\newblock \bibinfo{title}{Highly selective nanostructure sensors and methods of
  detecting target analytes} (\bibinfo{year}{2018}).
\newblock \bibinfo{note}{US Patent 9,983,183}.

\bibitem{silva2014performing}
\bibinfo{author}{Silva, A.} \emph{et~al.}
\newblock \bibinfo{journal}{\bibinfo{title}{Performing mathematical operations
  with metamaterials}}.
\newblock {\emph{\JournalTitle{Science}}} \textbf{\bibinfo{volume}{343}},
  \bibinfo{pages}{160--163} (\bibinfo{year}{2014}).

\bibitem{kojima2017acceleration}
\bibinfo{author}{Kojima, K.}, \bibinfo{author}{Wang, B.},
  \bibinfo{author}{Kamilov, U.}, \bibinfo{author}{Koike-{Akino}, T.} \&
  \bibinfo{author}{Parsons, K.}
\newblock \bibinfo{title}{Acceleration of {FDTD}-based inverse design using a
  neural network approach}.
\newblock In \emph{\bibinfo{booktitle}{Integrated Photonics Research, Silicon
  and Nanophotonics}}, \bibinfo{pages}{ITu1A--4}
  (\bibinfo{organization}{Optical Society of America}, \bibinfo{year}{2017}).

\bibitem{teng2018broadband}
\bibinfo{author}{Teng, M.} \emph{et~al.}
\newblock \bibinfo{title}{Broadband soi mode order converter based on topology
  optimization}.
\newblock In \emph{\bibinfo{booktitle}{2018 Optical Fiber Communications
  Conference and Exposition (OFC)}}, \bibinfo{pages}{1--3}
  (\bibinfo{year}{2018}).

\bibitem{lecun2015deep}
\bibinfo{author}{LeCun, Y.}, \bibinfo{author}{Bengio, Y.} \&
  \bibinfo{author}{Hinton, G.}
\newblock \bibinfo{journal}{\bibinfo{title}{Deep learning}}.
\newblock {\emph{\JournalTitle{Nature}}} \textbf{\bibinfo{volume}{521}},
  \bibinfo{pages}{436--444} (\bibinfo{year}{2015}).

\bibitem{krizhevsky2012imagenet}
\bibinfo{author}{Krizhevsky, A.}, \bibinfo{author}{Sutskever, I.} \&
  \bibinfo{author}{Hinton, G.~E.}
\newblock \bibinfo{title}{Imagenet classification with deep convolutional
  neural networks}.
\newblock vol.~\bibinfo{volume}{60}, \bibinfo{pages}{84--90},
  \doiprefix\url{10.1145/3065386} (\bibinfo{publisher}{ACM},
  \bibinfo{address}{New York, NY, USA}, \bibinfo{year}{2017}).

\bibitem{ghaboussi1991knowledge}
\bibinfo{author}{Ghaboussi, J.}, \bibinfo{author}{Garrett~Jr, J.} \&
  \bibinfo{author}{Wu, X.}
\newblock \bibinfo{journal}{\bibinfo{title}{Knowledge-based modeling of
  material behavior with neural networks}}.
\newblock {\emph{\JournalTitle{Journal of Engineering Mechanics}}}
  \textbf{\bibinfo{volume}{117}}, \bibinfo{pages}{132--153}
  (\bibinfo{year}{1991}).

\bibitem{baldi2014searching}
\bibinfo{author}{Baldi, P.}, \bibinfo{author}{Sadowski, P.} \&
  \bibinfo{author}{Whiteson, D.}
\newblock \bibinfo{journal}{\bibinfo{title}{Searching for exotic particles in
  high-energy physics with deep learning}}.
\newblock {\emph{\JournalTitle{Nature Communications}}}
  \textbf{\bibinfo{volume}{5}} (\bibinfo{year}{2014}).

\bibitem{jun2018deep}
\bibinfo{author}{Jun, Y.} \emph{et~al.}
\newblock \bibinfo{journal}{\bibinfo{title}{Deep-learned 3d black-blood imaging
  using automatic labelling technique and 3d convolutional neural networks for
  detecting metastatic brain tumors}}.
\newblock {\emph{\JournalTitle{Scientific reports}}}
  \textbf{\bibinfo{volume}{8}} (\bibinfo{year}{2018}).

\bibitem{radovic2018machine}
\bibinfo{author}{Radovic, A.} \emph{et~al.}
\newblock \bibinfo{journal}{\bibinfo{title}{Machine learning at the energy and
  intensity frontiers of particle physics}}.
\newblock {\emph{\JournalTitle{Nature}}} \textbf{\bibinfo{volume}{560}}
  (\bibinfo{year}{2018}).

\bibitem{liu2018training}
\bibinfo{author}{Liu, D.}, \bibinfo{author}{Tan, Y.}, \bibinfo{author}{Khoram,
  E.} \& \bibinfo{author}{Yu, Z.}
\newblock \bibinfo{journal}{\bibinfo{title}{Training deep neural networks for
  the inverse design of nanophotonic structures}}.
\newblock {\emph{\JournalTitle{ACS Photonics}}} \textbf{\bibinfo{volume}{5}},
  \bibinfo{pages}{1365--1369} (\bibinfo{year}{2018}).

\bibitem{ma2018deep}
\bibinfo{author}{Ma, W.}, \bibinfo{author}{Cheng, F.} \& \bibinfo{author}{Liu,
  Y.}
\newblock \bibinfo{journal}{\bibinfo{title}{Deep-learning enabled on-demand
  design of chiral metamaterials}}.
\newblock {\emph{\JournalTitle{ACS Nano}}} \textbf{\bibinfo{volume}{12}},
  \bibinfo{pages}{6326--6334} (\bibinfo{year}{2018}).

\bibitem{malkiel2018deep}
\bibinfo{author}{Malkiel, I.} \emph{et~al.}
\newblock \bibinfo{title}{Deep learning for the design of nano-photonic
  structures}.
\newblock In \emph{\bibinfo{booktitle}{2018 IEEE International Conference on
  Computational Photography (ICCP)}}, \bibinfo{pages}{1--14},
  \doiprefix\url{10.1109/ICCPHOT.2018.8368462} (\bibinfo{year}{2018}).

\bibitem{peurifoy2018nanophotonic}
\bibinfo{author}{Peurifoy, J.} \emph{et~al.}
\newblock \bibinfo{journal}{\bibinfo{title}{Nanophotonic particle simulation
  and inverse design using artificial neural networks}}.
\newblock {\emph{\JournalTitle{Science Advances}}}
  \textbf{\bibinfo{volume}{4}}, \doiprefix\url{10.1126/sciadv.aar4206}
  (\bibinfo{year}{2018}).
\newblock
  \eprint{http://advances.sciencemag.org/content/4/6/eaar4206.full.pdf}.

\bibitem{sun2018efficient}
\bibinfo{author}{Sun, Y.}, \bibinfo{author}{Xia, Z.} \&
  \bibinfo{author}{Kamilov, U.~S.}
\newblock \bibinfo{journal}{\bibinfo{title}{Efficient and accurate inversion of
  multiple scattering with deep learning}}.
\newblock {\emph{\JournalTitle{Optics Express}}} \textbf{\bibinfo{volume}{26}},
  \bibinfo{pages}{14678--14688} (\bibinfo{year}{2018}).

\bibitem{asano2018optimization}
\bibinfo{author}{Asano, T.} \& \bibinfo{author}{Noda, S.}
\newblock \bibinfo{journal}{\bibinfo{title}{Optimization of photonic crystal
  nanocavities based on deep learning}}.
\newblock {\emph{\JournalTitle{arXiv preprint arXiv:1808.05722}}}
  (\bibinfo{year}{2018}).

\bibitem{tait2017neuromorphic}
\bibinfo{author}{Tait, A.~N.} \emph{et~al.}
\newblock \bibinfo{journal}{\bibinfo{title}{Neuromorphic photonic networks
  using silicon photonic weight banks}}.
\newblock {\emph{\JournalTitle{Scientific Reports}}}
  \textbf{\bibinfo{volume}{7}} (\bibinfo{year}{2017}).

\bibitem{mehrabian2018pcnna}
\bibinfo{author}{Mehrabian, A.}, \bibinfo{author}{Al-Kabani, Y.},
  \bibinfo{author}{Sorger, V.~J.} \& \bibinfo{author}{El-Ghazawi, T.}
\newblock \bibinfo{journal}{\bibinfo{title}{Pcnna: A photonic convolutional
  neural network accelerator}}.
\newblock {\emph{\JournalTitle{arXiv preprint arXiv:1807.08792}}}
  (\bibinfo{year}{2018}).

\bibitem{lin2018all}
\bibinfo{author}{Lin, X.} \emph{et~al.}
\newblock \bibinfo{journal}{\bibinfo{title}{All-optical machine learning using
  diffractive deep neural networks}}.
\newblock {\emph{\JournalTitle{Science}}}
  \doiprefix\url{10.1126/science.aat8084} (\bibinfo{year}{2018}).
\newblock
  \eprint{http://science.sciencemag.org/content/early/2018/07/25/science.aat8084.full.pdf}.

\bibitem{chiles2018design}
\bibinfo{author}{Chiles, J.}, \bibinfo{author}{Buckley, S.~M.},
  \bibinfo{author}{Nam, S.~W.}, \bibinfo{author}{Mirin, R.~P.} \&
  \bibinfo{author}{Shainline, J.~M.}
\newblock \bibinfo{journal}{\bibinfo{title}{Design, fabrication, and metrology
  of 10$\times$ 100 multi-planar integrated photonic routing manifolds for
  neural networks}}.
\newblock {\emph{\JournalTitle{APL Photonics}}} \textbf{\bibinfo{volume}{3}}
  (\bibinfo{year}{2018}).

\bibitem{piggott2015inverse}
\bibinfo{author}{Piggott, A.~Y.} \emph{et~al.}
\newblock \bibinfo{journal}{\bibinfo{title}{Inverse design and demonstration of
  a compact and broadband on-chip wavelength demultiplexer}}.
\newblock {\emph{\JournalTitle{Nature Photonics}}}
  \textbf{\bibinfo{volume}{9}}, \bibinfo{pages}{374--377}
  (\bibinfo{year}{2015}).

\bibitem{piggott2017fabrication}
\bibinfo{author}{Piggott, A.~Y.}, \bibinfo{author}{Petykiewicz, J.},
  \bibinfo{author}{Su, L.} \& \bibinfo{author}{Vu{\v{c}}kovi{\'c}, J.}
\newblock \bibinfo{journal}{\bibinfo{title}{Fabrication-constrained
  nanophotonic inverse design}}.
\newblock {\emph{\JournalTitle{Scientific Reports}}}
  \textbf{\bibinfo{volume}{7}}, \bibinfo{pages}{1786} (\bibinfo{year}{2017}).

\bibitem{frandsen2016inverse}
\bibinfo{author}{Frandsen, L.~H.} \& \bibinfo{author}{Sigmund, O.}
\newblock \bibinfo{title}{Inverse design engineering of all-silicon
  polarization beam splitters}.
\newblock In \emph{\bibinfo{booktitle}{Photonic and Phononic Properties of
  Engineered Nanostructures VI}}, vol. \bibinfo{volume}{9756},
  \bibinfo{pages}{97560Y} (\bibinfo{organization}{International Society for
  Optics and Photonics}, \bibinfo{year}{2016}).

\bibitem{cao2003adjoint}
\bibinfo{author}{Cao, Y.}, \bibinfo{author}{Li, S.}, \bibinfo{author}{Petzold,
  L.} \& \bibinfo{author}{Serban, R.}
\newblock \bibinfo{journal}{\bibinfo{title}{Adjoint sensitivity analysis for
  differential-algebraic equations: The adjoint dae system and its numerical
  solution}}.
\newblock {\emph{\JournalTitle{SIAM Journal on Scientific Computing}}}
  \textbf{\bibinfo{volume}{24}}, \bibinfo{pages}{1076--1089}
  (\bibinfo{year}{2003}).

\bibitem{he2016deep}
\bibinfo{author}{He, K.}, \bibinfo{author}{Zhang, X.}, \bibinfo{author}{Ren,
  S.} \& \bibinfo{author}{Sun, J.}
\newblock \bibinfo{title}{Deep residual learning for image recognition}.
\newblock In \emph{\bibinfo{booktitle}{Proceedings of the IEEE conference on
  computer vision and pattern recognition}}, \bibinfo{pages}{770--778}
  (\bibinfo{year}{2016}).

\bibitem{kang2014resolution}
\bibinfo{author}{Kang, Z.} \emph{et~al.}
\newblock \bibinfo{journal}{\bibinfo{title}{Resolution-enhanced all-optical
  analog-to-digital converter employing cascade optical quantization
  operation}}.
\newblock {\emph{\JournalTitle{Optics Express}}} \textbf{\bibinfo{volume}{22}},
  \bibinfo{pages}{21441--21453} (\bibinfo{year}{2014}).

\bibitem{tian2018broadband}
\bibinfo{author}{Tian, Y.} \emph{et~al.}
\newblock \bibinfo{journal}{\bibinfo{title}{Broadband 1$\times$ 3 couplers with
  variable splitting ratio using cascaded step-size mmi}}.
\newblock {\emph{\JournalTitle{IEEE Photonics Journal}}}
  \textbf{\bibinfo{volume}{10}}, \bibinfo{pages}{1--8} (\bibinfo{year}{2018}).

\bibitem{xu2017integrated}
\bibinfo{author}{Xu, K.} \emph{et~al.}
\newblock \bibinfo{journal}{\bibinfo{title}{Integrated photonic power divider
  with arbitrary power ratios}}.
\newblock {\emph{\JournalTitle{Optics Letters}}} \textbf{\bibinfo{volume}{42}},
  \bibinfo{pages}{855--858} (\bibinfo{year}{2017}).

\bibitem{lu2016ultra}
\bibinfo{author}{Lu, L.}, \bibinfo{author}{Zhang, M.}, \bibinfo{author}{Zhou,
  F.} \& \bibinfo{author}{Liu, D.}
\newblock \bibinfo{title}{An ultra-compact colorless 50: 50 coupler based on
  {PhC}-like metamaterial structure}.
\newblock In \emph{\bibinfo{booktitle}{Optical Fiber Communications Conference
  and Exhibition (OFC), 2016}}, \bibinfo{pages}{1--3}
  (\bibinfo{organization}{IEEE}, \bibinfo{year}{2016}).

\bibitem{kingma2014adam}
\bibinfo{author}{Kingma, D.~P.} \& \bibinfo{author}{Ba, J.}
\newblock \bibinfo{journal}{\bibinfo{title}{Adam: A method for stochastic
  optimization}}.
\newblock {\emph{\JournalTitle{arXiv preprint arXiv:1412.6980}}}
  (\bibinfo{year}{2014}).

\bibitem{palik1997hand}
\bibinfo{author}{Palik, E.~D.}
\newblock \bibinfo{title}{Handbook of optical constants of solids}.
\newblock In \emph{\bibinfo{booktitle}{Handbook of optical constants of
  solids}}, \bibinfo{pages}{429--443} (\bibinfo{publisher}{Elsevier},
  \bibinfo{year}{1997}).

\end{thebibliography}

\section*{Author contributions statement}
M.T. and K.K. conceived the device idea(s) and numerical tests, M.T., D.J., and T.A.K. developed the neural network scripts, M.T. conducted the numerical and neural network tests and analyzed the results. All authors helped interpreting the results and reviewed the manuscript. 

\section*{Additional information}
\textbf{Competing interests}: Authors declare no competing financial interest. 
\end{document}